\documentclass[epj]{svjour}
\begin{document}
\title{The Aharonov-Casher effect for spin-1 particles in
non-commutative quantum mechanics }
\author{Sayipjamal Dulat\inst{1}\inst{3} \and Kang Li\inst{2}\inst{3}
}                     
%
%
\institute{ School of Physics Science and Technology, Xinjiang
University, Urumqi, 830046, China\and Department of Physics,
Hangzhou Normal University, Hangzhou, 310036, China\and The Abdus
Salam International Center for Theoretical Physics, Trieste, Italy}
\date{Received: date / Revised version: date}
%
\abstract{ By using a generalized Bopp's shift formulation, instead
of star product method, we investigate the Aharonov-Casher(AC)
effect for a spin-1 neutral particle in non-commutative(NC) quantum
mechanics. After solving the Kemmer equations both on a
non-commutative space and a non-commutative phase space, we obtain
the corrections to the topological phase of the AC effect for a
spin-1 neutral particle both on a NC space and a NC phase space.
\PACS{
      {PACS-key}{02.40.Gh, 11.10.Nx, 03.65.-w}
     } 
} 
\maketitle

\section{Introduction}
Recently, there has been an increasing interest in the study of
physics on a non-commutative space. Apart from non-commutative field
theories, there are many papers devoted to the study of various
aspects of quantum mechanics on a NC space with usual (commutative)
time coordinate \cite{AB-2}-\cite{kang}. For example, the
Aharonov-Bohm phase on a NC space and a NC phase space has been
studied in Refs. \cite{AB-2}-\cite{AB-4}. The Aharonov-Casher phase
for a spin-$1/2$ particle on a NC space and a NC phase space has
been studied in Refs.\cite{AC-NCS spin-half}-\cite{AC-NCPS
spin-half}. Some features of the AC effect for a spin-$1$ neutral
particle on a non-commutative space has been investigated in
Ref.\cite{AC-NCS spin-one} by using the star product formulation. It
is interesting to obtain the corrections to the topological phase of
the AC effect for a spin-1 neutral particle both on a NC space and a
NC phase space by using the new method  in Ref.\cite{AB-4}.

This article is organized as follows: in section \ref{AC-CS}, by
using the Lagrangian formulation, we discuss the AC effect on a
commutative space. In section \ref{AC-NCS}, we study the AC effect
on a NC space and give a generalized formula of the AC phase. In
section \ref{AC-NC-PS},
 a generalized formula of AC
phase on a NC phase space is given. The conclusions are presented in
the last section.
\section{AC effect for spin-1 particles on a commutative space
time}\label{AC-CS}
 In this section by following Ref.\cite{AC-QM
spin1} we review briefly the Aharonov-Casher effect of a spin-1
particle on a commutative space time.  The Lagrangian for a free
spin-1 particle of mass m is
\begin{equation}\label{free-action}
L = \bar \phi (i\beta^\nu \partial_\nu - m ) \phi \;,
\end{equation}
where the $10\times 10$  matrices $\beta^\nu$ are generalization of
the $4\times 4$ Dirac gamma matrices, and it can be chosen as
follows\cite{AC-QM spin1}-\cite{QFT spin1}

$$ \beta^0 =\left(\matrix{\widehat  O & \widehat O & I &  o^{\dag}\cr
                            \widehat O &\widehat O  & \widehat O & o^{\dag} \cr
                               I &\widehat O  & \widehat O & o^{\dag} \cr
                              o&o&o&0\cr} \right),
                              \\
 \beta^j=\left(\matrix{\widehat  O & \widehat  O  & \widehat  O  & -i {K}^{j \dag}\cr
                                 \widehat O & \widehat O  &S^j &  o^{\dag} \cr
                                 \widehat O &-S^j & \widehat O & o^{\dag} \cr
                                 -iK^j &o&o&0\cr}  \right),
                                 $$
\noindent with $j=1,2,3$. The elements of the $10\times 10$ matrices
$\beta^\nu$ are given by the matrices

$$ \widehat O = \left(\matrix{0 & 0  & 0  \cr
                                 0& 0  & 0  \cr
                                 0 & 0 & 0 \cr}  \right),
                                 \hspace{1cm}
                                I = \left(\matrix{1 & 0  & 0  \cr
                                                            0& 1  & 0  \cr
                                                            0 & 0 & 1 \cr}  \right), $$

$$S^1 = i  \left(\matrix{0 & 0  & 0  \cr
                                 0& 0  & -1  \cr
                                 0 & 1 & 0 \cr}  \right),
                                 S^2= i  \left(\matrix{0 & 0  & 1  \cr
                                 0& 0  & 0 \cr
                                 -1 & 0 & 0 \cr}  \right),
                                 S^3= i  \left(\matrix{0 & -1  & 0  \cr
                                 1& 0  & 0 \cr
                                 0 & 0 & 0 \cr} \right),$$

$$ o = \left(\matrix{0 & 0  & 0 \cr} \right), \hspace{1cm}  K^1 = \left(\matrix{1 & 0  & 0 \cr}
\right), $$
 $$  K^2 = \left(\matrix{0 & 1  & 0 \cr} \right), \hspace{1cm}
   K^3 = \left(\matrix{0 & 0  & 1 \cr} \right). $$

\noindent The above $\beta$ matrices satisfy the following
relation
\begin{equation}\label{beta-algebra}
\beta_\nu \beta_\lambda \beta_\rho + \beta_\rho \beta_\lambda
\beta_\nu =\beta_\nu g_{\lambda\rho} + \beta_\rho g_{\nu\lambda}.
\end{equation}

\noindent Other algebraic properties of the Kemmer $\beta$-matrices
were given in Ref.\cite{kemmer-theory}; the metric tensor is
$g_{\lambda\rho} = \mbox{diag}(1,-1,-1,-1)$. The Kemmer equation of
motion is
\begin{equation}\label{free-kemmer}
(i\beta^\nu \partial_\nu - m ) \phi =0 .
\end{equation}
The Lagrangian for  a spin-1 neutral particle with a magnetic
dipole moment $\mu_m$ interacting  with the electromagnetic field
has the form
\begin{eqnarray}\label{action-int}
L = \bar \phi( i \beta^\nu \partial_\nu + {1\over 2} \mu_m
S_{\lambda\rho} F^{\lambda\rho} - m)\phi  \;,
\end{eqnarray}
where $F^{\lambda\rho}$ is the field strength of the
electromagnetic field; $S_{\lambda\rho}$ is the Dirac
$\sigma_{\lambda\rho}$ like spin operator, which can be defined as
\begin{equation}\label{spin-s}
S_{\lambda\rho} = \frac{1}{2}(\beta_\lambda\beta_\rho
-\beta_\rho\beta_\lambda)\;.
\end{equation}
It follows that in the presence of an electromagnetic field, the
Kemmer equation of motion of a spin-1 neutral particle with a
 magnetic moment $\mu_m$ is
\begin{equation}\label{int-kemmer}
( i \beta^\nu \partial_\nu + {1\over 2} \mu_m S_{\lambda\rho}
F^{\lambda\rho} - m)\phi = 0 \;.
\end{equation}
The aim is to find a solution of the above equation, which can be
written in the following form
\begin{equation}\label{solution-free-kemmer}
\phi = e^{-i\xi_3 \int^r  {\mathbf A}^\prime \cdot \;d\mathbf
r}\;\phi_0,
\end{equation}
where $\phi_0$ is a solution of  (\ref{free-kemmer}); the spin-1
pseudo-vector operator $\xi_\nu$ in (\ref {solution-free-kemmer}) is
defined as
\begin{equation}\label{spin xi}
\xi_\nu
=\frac{1}{2}i\varepsilon_{\nu\lambda\rho\sigma}\beta^\lambda\beta^\rho\beta^\sigma\;,
\end{equation}
where $\varepsilon_{\nu\lambda\rho\sigma}$ is the Levi-Civita symbol
in four dimensions. Now we need to find the explicit form of the
vector ${\mathbf A}^\prime$ in (\ref{solution-free-kemmer}). To do
this, first we write Eq.(\ref{free-kemmer}) for $\phi_0$ in terms of
$\phi$
\begin{equation}\label{free-kemmer-2}
( i \beta^\nu \partial_\nu- m)\;e^{i\xi_3 \int^r {\mathbf A}^\prime
\cdot \;d{\mathbf r}}\;\phi= 0.
\end{equation}
Then the equivalence  of (\ref{int-kemmer}) and
(\ref{free-kemmer-2}) can be obtained by imposing the following two
conditions
\begin{equation}\label{condition-1}
e^{-i\xi_3 \int^r {\mathbf A}^\prime \cdot \;d{\mathbf
r}}\;\beta^\nu\;e^{i\xi_3 \int^r {\mathbf A}^\prime \cdot
\;d{\mathbf r}}= \beta^\nu,
\end{equation}
and
\begin{equation}\label{condition-2}
-\beta^\nu\xi_3 A^\prime_\nu \phi ={1\over 2} \mu_m
S_{\lambda\rho} F^{\lambda\rho}\phi = \mu_m S_{0l} F^{0l}\phi.
\end{equation}
By  comparing  Eq. (\ref{condition-1}) with the Baker-Housdorf
formula
\begin{eqnarray}\label{housdorf}
e^{-i\lambda\xi_3}\;\beta^\nu \;e^{i\lambda\xi_3} &=& \beta^\nu +
\wp(-i\lambda)[\xi_3,\beta^\nu] \nonumber\\ & + &
\frac{1}{2!}\wp(-i\lambda)^2[\xi_3,[\xi_3,\beta^\nu]]\ldots,
\end{eqnarray}
one obtains, $ [\xi_3,\beta^\nu]=0$;  $\wp$ in (\ref{housdorf})
stands for the path ordering of the integral in the phase. If $\nu
\neq 3$ this commutation relation is automatically satisfied. For
$\nu=3$, by using (\ref{beta-algebra}) and (\ref{spin xi}), one
finds that the commutator does not vanish. Therefore in order to
fulfil the first condition the particle is restricted to move in
$x-y$ plane, that is, $p_z=0$. In particular $\partial_3\phi=0$ and
$\hat A^\prime_3=0$. From second condition (\ref{condition-2}), by
using (\ref{beta-algebra}), (\ref{spin-s}) and (\ref{spin xi}), one
obtains
\begin{equation}
 A^\prime_1 = -2\mu_m E_2 \;,\hspace{2cm}  A^\prime_2 =
2\mu_m E_1 \;.
\end{equation}
Thus the AC phase for a neutral spin-1 particle moving in a $2+1$
space time under the influence of a pure electric field produced
by a uniformly charged infinitely long filament perpendicular to
the plane is
\begin{eqnarray}\label{AC-phase}
\phi_{AC} = \xi_3\oint {\mathbf A}^\prime \cdot\; d{\mathbf r} & = &
2\mu_m \xi_3\oint (E_1 dx_2 - E_2 dx_1) \nonumber\\ &=& 2\mu_m
\xi_3\varepsilon^{lk}\oint E_ldx_k\;.
\end{eqnarray}
 The above equation can also be written as in Ref.\cite{AC-QM
spin1}
\begin{eqnarray}\label{AC-phase-1}
\phi_{AC} = \xi_3\oint {\mathbf A}^\prime \cdot\; d{\mathbf r}=
\xi_3\int_S({\mathbf\nabla}\times {\mathbf A}^\prime)\cdot d{\mathbf
S}\nonumber\\ = 2\mu_m\xi_3 \int_S ({\mathbf \nabla}\cdot{\mathbf
E}) dS = 2\mu_m\xi_3 \lambda_e\;,
\end{eqnarray}
where $\lambda_e$ is the charge density of the filament. This spin-1
AC phase is a purely quantum mechanical effect and has no classical
interpretation.   One may note that the AC phase for spin-1
particles is exactly the same as in the case of spin-$\frac{1}{2}$,
except that the spin and spinor have changed. The factor of two
shows that the phase is twice that accumulated by a
spin-$\frac{1}{2}$ particle with the same magnetic dipole moment
coupling constant, in the same electric field.
\section{ AC effect for spin-1 particles on a non-commutative
space}\label{AC-NCS} On a NC space the coordinate and momentum
operators satisfy the following commutation relations (we take
$\hbar=c=1$ unit)
\begin{equation}\label{eq1}
~[\hat{x}_{i},\hat{x}_{j}]=i\Theta_{ij},~~~
[\hat{p}_{i},\hat{p}_{j}]=0,~~~[\hat{x}_{i},\hat{p}_{j}]=i
\delta_{ij},
\end{equation}
where $\Theta_{ij}$ is an element of an antisymmetric matrix, it is
related to the energy scale and it represents the non-commutativity
of the NC space; $\hat{x}_i$ and $\hat{p}_i$ are the coordinate and
momentum operators on  a NC space.

By replacing the usual product in (\ref{int-kemmer}) with a star
product (Moyal-Weyl product), the Kemmer equation
 for a spin-1 neutral particle with a magnetic
dipole moment $\mu_m$, on the NC space, can be written as

\begin{equation}\label{int-kemmer-NC-1}
( i \beta^\nu \partial_\nu + {1\over 2} \mu_m S_{\lambda\rho}
F^{\lambda\rho} - m)\ast\phi = 0.
\end{equation}
The star product between two functions is defined by,
\begin{eqnarray}\label{eq4}
(f  \ast g)(x) &=& e^{  \frac{i}{2}
 \Theta_{ij} \partial_{x_i} \partial_{x_j}
 }f(x_i)g(x_j)   \nonumber\\  &=&  f(x)g(x)
 + \frac{i}{2}\Theta_{ij} \partial_i f \partial_j
 g\big|_{x_i=x_j}+{\mathcal{O}}(\Theta^2).\nonumber\\
\end{eqnarray}
here $f(x)$ and $g(x)$ are two arbitrary functions.

On a NC space the star product can be changed into an ordinary
product by a Bopp's shift, that is, by shifting coordinates $x_\nu$
with
\begin{equation}\label{eq19}
x_\nu \rightarrow
\hat{x}_\nu=x_\nu-\frac{1}{2}\Theta_{\nu\lambda}p^\lambda.
\end{equation}
Now, let us consider the non-commutative Kemmer equation
(\ref{int-kemmer-NC-1}). To replace the star product in
(\ref{int-kemmer-NC-1}) with an ordinary product, the
$F_{\nu\lambda}$ must, up to the first order of the NC parameter
$\Theta$, be shifted \cite{AB-4} as
\begin{equation}\label{eq21}
F_{\nu\lambda}\rightarrow\hat{F}_{\nu\lambda}=F_{\nu\lambda}+\frac{1}{2}
 \Theta^{\rho\sigma} p_\rho \partial_\sigma F_{\nu\lambda}.
\end{equation}
which is equivalent to the  Bopp's shift (\ref{eq19}). Then the
Kemmer equation on a NC space has the form
\begin{equation}\label{int-kemmer-NC}
( i \beta^\nu \partial_\nu + {1\over 2} \mu_m S_{\lambda\rho}
\hat{F}^{\lambda\rho} - m)\phi = 0.
\end{equation}
In a similar way as the commuting space, the solution of the above
equation can also be written as
\begin{equation}\label{phi-NC}
\phi = e^{-i\xi_3 \int^r \hat{\mathbf A}^\prime \cdot \;d\mathbf
r}\;\phi_0.
\end{equation}
To determine $\hat {\mathbf A}^\prime$ we write the free Kemmer
equation as
\begin{equation}\label{free-kemmer-NC-2}
( i \beta^\nu \partial_\nu- m)\;e^{i\xi_3 \int^r \hat{\mathbf
A}^\prime \cdot \;d\mathbf r}\;\phi= 0.
\end{equation}
The equivalence of (\ref{int-kemmer-NC}) and
(\ref{free-kemmer-NC-2}) gives the following two conditions
\begin{equation}\label{condition-1-NC}
e^{-i\xi_3 \int^r \hat{\mathbf A}^\prime \cdot \;d\mathbf
r}\;\beta^\nu\;e^{i\xi_3 \int^r \hat{\mathbf r}^\prime \cdot
\;d\mathbf r}= \beta^\nu
\end{equation}
and
\begin{equation}\label{condition-2-NC}
-\beta^\nu\xi_3 \hat A^\prime_\nu \phi ={1\over 2} \mu_m
S_{\lambda\rho} \hat F^{\lambda\rho}\phi = \mu_m S_{0l}
\hat{F}^{0l}\phi.
\end{equation}
By using the Baker-Housdorf formula (\ref{housdorf}), the first
condition (\ref{condition-1-NC}) implies that, $
[\xi_3,\beta^\nu]=0$. If $\nu \neq 3$ then this commutation relation
is automatically satisfied. For $\nu=3$, by using
(\ref{beta-algebra}) and (\ref{spin xi}), one finds that the
commutator does not vanish. Therefore in order to fulfil the first
condition we restrict ourselves to $2+1$ space-time. In particular
$\partial_3\phi=0$ and $\hat A^\prime_3=0$. From second condition
(\ref{condition-2-NC}), by using (\ref{beta-algebra}),
(\ref{spin-s}) and (\ref{spin xi}), one obtains
\begin{eqnarray}
\hat A^\prime_1 &= &-2\mu_m \hat F^{02}= -2\mu_m F^{02} - 2\mu_m
\frac{1}{2}\Theta^{ij}p_i\partial_j F^{02} \nonumber\\
~&=&-2\mu_m E_2 -\mu_m\theta \varepsilon^{ij} p_i\partial_j E_2\;, \nonumber\\
\hat A^\prime_2 &=& 2\mu_m \hat F^{01}= 2\mu_m F^{01} + 2\mu_m
\frac{1}{2}\Theta^{ij}p_i\partial_j F^{01}  \\
~&=& 2\mu_m E_1 -\mu_m \theta \varepsilon^{ij} p_i\partial_j E_1
\nonumber
\end{eqnarray}
with $\Theta^{ij}=\theta\epsilon^{ij}$, $\Theta^{0\mu}=\Theta^{\mu
0}=0$; $\epsilon^{ij}=-\epsilon^{ji}$, $\epsilon^{12}=+1$. Thus the
AC phase for a neutral spin-1 particle moving in a $2+1$
non-commutative space under the influence of a pure electric field
produced by a uniformly charged infinitely long filament
perpendicular to the plane is
\begin{eqnarray}\label{AC-phase-NC}
\hat\phi_{AC} &=& \xi_3\oint \hat{\mathbf A}^\prime \cdot\;
d{\mathbf r}=2\mu_m \xi_3\varepsilon^{lk}\oint E_ldx_k \nonumber\\
& + & \mu_m \xi_3\theta \varepsilon^{ij}\varepsilon^{lk}\oint
p_i\partial_j E_l dx_k\;.
\end{eqnarray}
In a similar way as in  spin-$\frac{1}{2}$ \cite{AB-4} \cite{AC-NCPS
spin-half}, the momentum on a NC space for a spin-$1$ neutral
particle can also be written as
\begin{equation}\label{eq29}
p_i=mv_i+(\vec E\times\vec \mu)_i+\mathcal{O}(\theta),
\end{equation}
where $\vec{\mu}=2\mu_m \vec S$, and $\vec S$ is the spin operator
of the spin-1. By inserting (\ref{eq29}) into (\ref{AC-phase-NC}),
we have
\begin{equation}\label{eq31}
\hat{\phi}_{AC}=\phi_{AC}+\delta\phi_{NCS},
\end{equation}
where $\phi_{AC}$ is the AC phase in (\ref{AC-phase}) on a commuting
space; the  additional phase $\delta \phi_{NCS}$, related to the
non-commutativity of space, is given by
\begin{eqnarray}\label{eq32}
\delta\phi_{NCS}=\mu_m \xi_3\theta
\varepsilon^{ij}\varepsilon^{lk}\oint [k_i - (\vec\mu \times \vec
E)_i]\partial_j E_l dx_k\;,
\end{eqnarray}
where $k_i=mv_i$ is the wave number; $\xi_3$  represents  spin
degrees of freedom. If the spin of the neutral particle along the
$z$ direction, namely, $\vec{\mu}=2 \mu_m s_3 \hat{\vec{k}}$,
$\hat{\vec{k}}$ is a unite vector in the z direction, then our
results here are the same as
 the result of Ref. \cite{AC-NCS spin-one}, where the star product
calculation has been used.

\section{AC effect for spin-1 particles on a non-commutative phase
space}\label{AC-NC-PS}

In  section \ref{AC-NCS} we have investigated the AC effect for a
neutral spin-1 particle on a NC space, where space-momentum, and
space-space are non-commuting, but momentum-momentum are commuting.
The Bose-Einstein statistics in non-commutative quantum mechanics
requires both space-space and momentum-momentum non-commutating. The
NC space with non-commuting momentum-momentum  is called NC phase
space.  In this section we study the AC phase on a NC phase space.
On a NC phase space, the commutation relation in (\ref{eq1}) should
be replaced by
\begin{equation}
[\hat{p}_{i},\hat{p}_{j}]=i\bar{\Theta}_{ij},
\end{equation}
where $\bar{\Theta}$ is the antisymmetric matrix, its elements
represent the non-commutative property of the momenta.  Then  the
Kemmer equation for AC problem on a NC phase space has the form
\begin{equation}\label{int-kemmer-NCPS}
( -\beta^\nu p_\nu + {1\over 2} \mu_m S_{\lambda\rho}
F^{\lambda\rho} - m)\ast\phi = 0.
\end{equation}
 The star product in
(\ref{int-kemmer-NCPS}) on a NC phase space can be replaced by the
usual product in two steps, first we need to replace $x_i$ and $
p_i$ by a generalized Bopp's shift as
\begin{eqnarray}\label{gbshift1}
x_\nu\rightarrow {\hat x}_\nu = \alpha x_{\nu}-\frac{1}{2 \alpha}\Theta_{\nu\lambda}p^{\lambda},\nonumber\\
p_\nu\rightarrow {\hat p}_\nu = \alpha
p_\nu+\frac{1}{2\alpha}\bar{\Theta}_{\nu\lambda}x^{\lambda},
\end{eqnarray}
where $\alpha$ is the scaling parameter, and it is related to the
non-commutativity of the phase space via $\Theta\bar \Theta =
4\alpha^2(\alpha^2 -1 )\cdot {\bf I}$, here ${\bf I}$ is a unit
matrix. Then we also need to rewrite the shift in (\ref{eq21}) as
\begin{equation}\label{LW-shift}
F_{\nu\lambda}\rightarrow\hat{{\mathcal F}}_{\nu\lambda}=\alpha
F_{\nu\lambda}+\frac{1}{2\alpha}
 \Theta^{\rho\sigma} p_\rho\partial_\sigma F_{\nu\lambda}.
\end{equation}
Thus the Kemmer equation for AC problem on a NC phase space has the
form
\begin{equation}\label{int-kemmer-NCPS-2}
( - \beta^\nu \hat{p}_\nu
 +
{1\over 2} \mu_m S_{\lambda\rho} \hat {\mathcal F}^{\lambda\rho} -
m )\phi = 0.
\end{equation}
Since $\alpha\neq 0$, the above  equation can be written as
\begin{equation}\label{int-kemmer-NCPS-3}
\begin{array}{ll}
\Big( - \beta^\nu p_\nu -
\frac{1}{2\alpha^2}\beta^\nu\bar{\Theta}_{\nu\lambda}x^{\lambda} +
{1\over 2} \mu_m S_{\lambda\rho}(
F^{\lambda\rho}&~ \\
~~~~~~~~~~~+\frac{1}{2\alpha^2}
 \Theta^{\sigma\tau} p_\sigma\partial_\tau F^{\lambda\rho} ) -
m^\prime \Big)\phi = 0.&~
\end{array}
\end{equation}
where $m'=m/\alpha$. We write the above equation  in the following
form
\begin{equation}\label{free-kemmer-NCPS}
( - \beta^\nu p_\nu  - m^\prime)\;e^{\frac{i}{2\alpha^2 } \int^r
\bar{\Theta}_{\nu\lambda} x^{\lambda}dx^\nu  + i\xi_3 \int^r
\hat{\mathbf \mathcal A}^\prime \cdot \;d\mathbf{r}}\;\phi= 0 \;.
\end{equation}
To have the equivalence  of (\ref{int-kemmer-NCPS-3}) and
(\ref{free-kemmer-NCPS}), we impose the following two conditions

\begin{equation}\label{condition-1-NCPS}
e^{-i\xi_3 \int^r  \hat{\mathbf\mathcal A }^\prime \cdot
\;d{\mathbf r}}\;\beta^\nu\;e^{i\xi_3 \int^r \hat{\mathbf\mathcal
A}^\prime\cdot \;d{\mathbf r}}= \beta^\nu \;,
\end{equation}
and
\begin{equation}\label{condition-2-NCPS}
-\beta^\nu \xi_3 \hat{\mathcal A }^\prime_\nu \phi ={1\over
2\alpha} \mu_m S_{\lambda\rho} \hat F^{\lambda\rho}\phi =
\frac{\mu_m}{\alpha} S_{0l} \hat{F}^{0l}\phi \;.
\end{equation}
In an analogous way as in NC space, from (\ref{condition-1-NCPS})
and (\ref{condition-2-NCPS}) one may obtain
\begin{eqnarray}
\hat {\mathcal A }^\prime_1 &=&-\frac{2\mu_m}{\alpha} \hat F^{02}=
-2\mu_m F^{02} - 2\mu_m \frac{1}{2\alpha^2}\Theta^{ij}p_i\partial_j
F^{02}\nonumber\\~&=&-2\mu_m E_2 - \frac{\mu_m
\theta }{\alpha^2}\varepsilon^{ij} p_i\partial_j E_2\;,\nonumber\\
\hat {\mathcal A}^\prime_2 & =&\frac{2\mu_m}{\alpha} \hat F^{01}=
2\mu_m F^{01} + 2\mu_m \frac{1}{2\alpha^2}\Theta^{ij}p_i\partial_j
F^{01}\\~&=& 2\mu_m E_1 + \frac{\mu_m \theta }{\alpha^2} \theta
\varepsilon^{ij} p_i\partial_j E_1\;, \nonumber \\
\hat {\mathcal A}^\prime_3 & = & 0\;.\nonumber
\end{eqnarray}
Thus the AC phase for a neutral spin-1 particle moving in a $2+1$
non-commutative phase space  under the influence of a pure
electric field produced by a uniformly charged infinitely long
filament perpendicular to the plane is given by
\begin{eqnarray}\label{eq24}
\hat{\varphi}_{AC} &=& \frac{1}{2\alpha^2 } \oint
\bar{\Theta}_{\nu\lambda} x^{\lambda}dx^\nu  + \xi_3\oint
\hat{{\mathbf \mathcal A}}^\prime \cdot\; d{\mathbf r}\nonumber\\&
=& \frac{\theta}{2\alpha^2 } \oint \varepsilon^{ij} x_j dx_i +
2\mu_m \xi_3\varepsilon^{lk}\oint E_ldx_k \nonumber\\&+& \mu_m
\xi_3\frac{\theta }{\alpha^2}\varepsilon^{ij}\varepsilon^{lk}\oint
p_i\partial_j E_l dx_k
\end{eqnarray}
By $p_i=k^\prime_i+(\vec E\times\vec \mu)_i+\mathcal{O}(\theta)$,
and $k^\prime_i=m^\prime_i v_i$, $\vec{\mu}=2\mu_m \vec S$, one
obatains
\begin{equation}\label{AC-NCPS}
\hat{\varphi}_{AC} = \phi_{AC}+\delta\phi_{NCS}+\delta\phi_{NCPS},
\end{equation}
where $\phi_{AC}$  is the AC phase in (\ref{AC-phase}) on a
commuting space; $\delta\phi_{NCS}$ is the space-space non-commuting
contribution to the AC phase in (\ref{AC-phase}) , and its explicit
form is given in (\ref{eq32}); the last term $\delta\phi_{NCPS}$ is
the momentum-momentum non-commuting contribution to the AC phase in
(\ref{AC-phase}), and it has the form
\begin{eqnarray}\label{AC-NCPS-correction}
\delta\phi_{NCPS} &=& \frac{\bar{\theta}}{2\alpha^2 } \oint
\varepsilon^{ij} x_j dx_i \nonumber\\ & + & (\frac{1}{\alpha^2}
-1)\mu_m \xi_3\theta \varepsilon^{ij}\varepsilon^{lk}\oint
[k^\prime_i - (\vec\mu \times \vec E)_i]\partial_j E_l dx_k
\nonumber\\
\end{eqnarray}
which represents the non-commutativity of the momenta. The first
term in (\ref{AC-NCPS-correction}) comes from the momentum-momentum
non-commutativity; the second term is a velocity dependent
correction and does not have the topological properties of the
commutative AC effect and could modify the phase shift; the third
term is a correction to the vortex and does not contribute to the
line spectrum. In two dimensional non-commutative plane,
$\bar{\Theta}_{ij}=\bar{\theta}\epsilon_{ij}$, and the two NC
parameters $\theta$ and $\bar{\theta}$ are related by
$\bar{\theta}=4\alpha^2 (1-\alpha^2)/\theta$ \cite{kang}. When
$\alpha=1$, which leads to $\bar{\theta}_{ij}=0$,  the AC phase on a
NC phase space case reduces to the AC phase on a NC space case, i.e.
$\delta\phi_{NCPS}=0$ and equation (\ref{AC-NCPS}) changes into
equation (\ref{eq31}).

\section{Conclusions}

In this paper in order to study  the AC effect both on a
non-commutative space and on a non-commutative phase space, we use
the shift method, instead of star product formulation. Our shift
method is equivalent to the star product method, i.e., the Kemmer
equation with star product can be replaced by Bopp's shift together
with the shift which  we defined in (\ref{eq21}) for a NC space and
in (\ref{LW-shift}) for a NC phase space. The additional AC phase in
(\ref{eq31}) on a NC space is the same as the result of
Ref.\cite{AC-NCS spin-one}, where the star product method has been
used. Furthermore, by considering the momentum-momentum
non-commutativity we obtained the NC phase space corrections to
 the topological phase of the AC effect for a
spin-1 neutral particle. We note that the corrections (\ref{eq32})
and (\ref{AC-NCPS-correction}) to the topological phase
(\ref{AC-phase}) or (\ref{AC-phase-1}) of the AC effect for a spin-1
neutral particle both on a NC space  and  a NC phase space can be
obtained  from spin-$\frac{1}{2}$ corrections through the
replacement $\frac{1}{2}\gamma^0\sigma^{12}\longrightarrow \xi_3$.
One may conclude that, apart from the spin operators, the AC phase
for a higher spin  neutral particle is the same as in the case of
spin-$\frac{1}{2}$ and spin-1 in non-commutative quantum mechanics.

\section{Acknowledgments} Kang Li would like to thank Prof. S.
Randjbar-Daemi for his kind invitation and warm hospitality during
his visit at the ICTP. This work is supported in part by the
National Natural Science Foundation of China (10465004, 10665001 and
10575026) and Natural Science of Zhejiang Province (Y607437). The
authors also grateful to the support from the Abdus Salam ICTP,
Trieste, Italy.

\end{document}